\documentclass[aps,prd,reprint,preprintnumbers,showpacs,floatfix,nofootinbib,superscriptaddress]{revtex4-1}
\usepackage{amsmath}
\usepackage{amssymb}
\usepackage{latexsym}
\usepackage{graphics}
\usepackage{graphicx}
\usepackage{slashed}
\usepackage{color}
\usepackage{listings}
\usepackage{bm,bbm}
\usepackage[bookmarks=true,bookmarksopen=true,bookmarksnumbered=true,bookmarksopenlevel=3]{hyperref}

\definecolor{codegreen}{rgb}{0,0.6,0}
\definecolor{codegray}{rgb}{0.5,0.5,0.5}
\definecolor{codepurple}{rgb}{0.58,0,0.82}
\definecolor{backcolour}{rgb}{0.95,0.95,0.92}

\begin{document}

\title{Momentum anisotropy effects for quarkonium in a weakly-coupled quark-gluon plasma below the melting temperature}

\author{S.~Biondini}
\affiliation{Albert Einstein Center, Institute for Theoretical Physics, University of Bern, Sidlerstrasse 5, CH-3012 Bern, Switzerland}
\author{N.~Brambilla}
\affiliation{Physik-Department, Technische Universit\"{a}t M\"{u}nchen, James-Franck-Str. 1, 85748 Garching, Germany}
\affiliation{Institute for Advanced Study, Technische Universit\"{a}t M\"{u}nchen,\\ Lichtenbergstrasse 2 a, 85748 Garching, Germany}
\author{M.~A.~Escobedo}
\affiliation{Department of Physics, P.O. Box 35, 40014 University of Jyv\"askyl\"a, Finland}
\affiliation{Institut de Physique Th\'eorique, Universit\'e Paris Saclay, CNRS, CEA, F-91191 Gif-sur-Yvette, France}
\author{A.~Vairo}
\affiliation{Physik-Department, Technische Universit\"{a}t M\"{u}nchen, James-Franck-Str. 1, 85748 Garching, Germany}

\date{\today}

\preprint{TUM-EFT 88/16}

\begin{abstract}
In the early stages of heavy-ion collisions, the hot QCD matter expands more longitudinally than transversely.
This imbalance causes the system to become rapidly colder in the longitudinal direction and a local momentum anisotropy appears. 
In this paper, we study the heavy-quarkonium spectrum in the presence of a small plasma anisotropy. 
We work in the framework of pNRQCD at finite temperature.
We inspect arrangements of non-relativistic and thermal scales complementary to those considered in the literature. 
In particular, we consider temperatures larger and Debye masses smaller than the binding energy, 
which is a temperature range relevant for presently running LHC experiments.
In this setting we compute the leading thermal corrections to the binding energy and the thermal width induced by quarkonium gluo-dissociation.
\end{abstract}

\maketitle

\section{Introduction}
In present day experiments at the Large Hadron Collider (LHC) and at the Relativistic Heavy Ion Collider (RHIC) 
a rich and broad program is ongoing to investigate QCD at finite temperature. 
The establishment of a hot QCD medium, dubbed as quark-gluon plasma (QGP), has been inferred thanks to the observation of at least 
two striking signatures: jet quenching and quarkonia suppression. 
In particular the latter, which has been proposed since long as a probe of the QGP formation~\cite{Matsui:1986dk}, 
will be the subject of the present investigation.

Together with the experimental activity also the theoretical understanding of heavy quarkonia in medium has progressed significantly in the last years. 
A key to it has been the study of the heavy quark-antiquark potential in a thermal environment.
The  heavy quark-antiquark potential has been derived at high temperatures 
($T \gg 1/r \gtrsim m_D$, where $r$ is the quark-antiquark distance and $m_D$ the Debye mass) 
in~\cite{Laine:2006ns,Laine:2007gj,Burnier:2007qm}, 
and further computed in a wider range of temperatures in an effective field theory framework of QCD 
in the static limit~\cite{Brambilla:2008cx} and for a large but finite heavy-quark mass~\cite{Brambilla:2010vq}.  
The real part of the potential shows at high temperatures Debye screening, which is a source of quarkonium dissociation.
The potential has also an imaginary part that stems from two further dissociation mechanisms: 
Landau damping~\cite{Grandchamp:2001pf,Grandchamp:2002wp} and gluo-dissociation~\cite{Kharzeev:1994pz,Xu:1995eb}.

A complete understanding of quarkonium in medium has to account for realistic QGP features. 
Among these is the momentum anisotropy of the thermal medium constituents.
Indeed highly Lorentz contracted nuclei collide along the beam-axis, 
so that the longitudinal expansion of the hot QCD medium is more important than the radial expansion perpendicular to the beam axis
(see, e.g.,~\cite{Strickland:2014pga}). 
At weak coupling this longitudinal expansion causes the system to quickly become much colder in the longitudinal than in the transverse direction, 
moreover the anisotropy can persist for a long time~\cite{Israel:1976tn,Baym:1984np,Martinez:2009mf,Florkowski:2010cf,Martinez:2010sc}. 
Recently the properties of an anisotropic QGP have been the subject of several investigations carried out in the framework of
viscous hydrodynamics~\cite{Romatschke:2003ms,Romatschke:2004jh,Mrowczynski:2004kv,Nopoush:2016qas,Nopoush:2014pfa}.

So far the effect of a local anisotropy on a quark-antiquark bound state has been taken into account 
via hard thermal loop (HTL) resummation of the gluon self energy, where a finite momentum
anisotropy is assigned to the degrees of freedom entering the loops~\cite{Dumitru:2007hy,Dumitru:2009fy,Burnier:2009yu}. 
Numerical solutions of the Schr\"odinger equation for the bound state show that the anisotropy tends to decrease the
effect of Landau damping and thus to increase the quarkonium melting temperature~\cite{Dumitru:2009ni,Margotta:2011ta}, 
whereas analytical estimates are found in~\cite{Burnier:2009yu}.
 
In this work, we assume the quarkonium to be a Coulombic system,  
so that its inverse size scales like $m \alpha_{\rm s}$, and its typical binding energy like $m \alpha_{\rm s}^2$, 
where $m$ and $\alpha_{\rm s}$ are the heavy-quark mass and strong coupling respectively. 
This is realized when  $m \alpha_{\rm s}$ is much larger than the temperature scale (moreover we consider negligible the effects of the hadronic scale $\Lambda_{\rm QCD}$). 
In particular, we aim at investigating the heavy-quarkonium spectrum when the relevant scales, the non-relativistic and thermal ones, satisfy the following hierarchy
\begin{equation}
m \gg m \alpha_{\rm s} \gg \pi T \gg m \alpha_{\rm s}^2 \gg m_D, \Lambda_{\rm QCD} \, ,
\label{hiera}
\end{equation}
and in the presence of a finite momentum anisotropy of the QGP constituents. 
In a weakly-coupled QGP, the Debye mass, $m_D$, scales like $m_D \sim gT$ and provides the inverse of an electric screening length. 
The hierarchy of scales \eqref{hiera} may be relevant for the $\Upsilon(1S)$, whose mass, inverse radius 
and binding energy are respectively $m \approx 5$~GeV, $ m \alpha_{\rm s} \approx 1.5$~GeV and $m \alpha_{\rm s}^2 \approx 0.5$~GeV~\cite{Vairo:2010bm}.  
In an expanding and then cooling QGP, the regime \eqref{hiera} is met at some point, say for $T \lesssim 2 T_c \approx 0.3$~GeV for bottomonium.  
Note that this temperature is below the bottomonium melting temperature~\cite{Escobedo:2008sy}.
In so doing we partly generalize the study carried out in~\cite{Brambilla:2010vq} for the isotropic case. 

Since the quarkonium is assumed to be a Coulombic system,  
we do not include in the real part of the potential any term to model a (screened) long-range interaction (as done, e.g., in~\cite{Margotta:2011ta}).
Such an inclusion would not be supported by the hierarchy of energy scales \eqref{hiera}.
The spectrum has also an imaginary part that provides the quarkonium width.
In the situation of interest for this work, $m \alpha_{\rm s}^2 \gg m_D$, gluo-dissociation is the dominant mechanism producing the thermal width.
Such a mechanism has been reinterpreted as and connected to the singlet-to-octet break up in potential non-relativistic QCD (pNRQCD) 
at finite temperature in~\cite{Brambilla:2011sg}.  

Following a common choice in the literature we implement a momentum anisotropy via distribution functions (B for Bose--Einstein, F for Fermi--Dirac) 
that read~\cite{Romatschke:2003ms,Romatschke:2004jh}
\begin{equation}
f^{\rm B,F}(\bm{q},\xi) \equiv N(\xi)\,f^{\rm B,F}_{\hbox{\tiny}\rm iso} \left( \sqrt{\bm{q}^2+\xi\,(\bm{q} \cdot \bm{n})^2} \right)  \, ,
\label{ani_distri_def}
\end{equation}
where $\xi$ is the anisotropy parameter, $N(\xi)=\sqrt{1+\xi}$ is a normalization factor that guarantees the same number of particles for the anisotropic and isotropic distribution functions
and $f^{\rm B,F}_{\hbox{\tiny}\rm iso}(q)$ is understood to be either a Bose--Einstein or a Fermi--Dirac isotropic distribution for gluons and quarks respectively. 
Hence $f^{\rm B,F}(\bm{q},\xi)$ is obtained from the corresponding isotropic distribution 
by removing particles with a large momentum component along the anisotropy direction $\bm{n}$, and accordingly $\xi>0$ parameterizes the anisotropy strength.  
The normalization factor $N(\xi)$ is often put to one in the literature, though its origin and impact have been discussed in~\cite{Philipsen:2009wg}. 
As far as the present work is concerned, we keep the normalization factor in the following calculations.
    
The outline of the paper is the following: 
in Sec.~\ref{sec_1} we compute the thermal modification of pNRQCD, pNRQCD$_{\rm{{HTL}}}$,  by integrating out the scale $\pi T$ in the presence of a momentum anisotropy.
At this stage and at our accuracy thermal effects are encoded in the singlet potential. 
In Sec.~\ref{sec_2} we compute in pNRQCD$_{\rm{{HTL}}}$ the temperature-dependent real and imaginary parts of the quarkonium spectrum. 
The latter corresponds to the quarkonium thermal width. 
Conclusion and discussion are found in Sec.~\ref{sec_3}.

\section{Matching pNRQCD to pNRQCD$_{\rm{{HTL}}}$}
\label{sec_1}
According to \eqref{hiera}, one has to integrate out the heavy-quark mass 
and the typical momentum transfer before dealing with any thermal effect. 
Hence our starting point is pNRQCD, whose coefficients can be obtained at zero temperature.  
The corresponding Lagrangian density reads as follows (we show only terms relevant for the present work)~\cite{Pineda:1997bj,Brambilla:1999xf,Brambilla:2004jw}:
\begin{eqnarray}
&&\mathcal{L}_{\rm{{pNRQCD}}}=-\frac{1}{4} F^a_{\mu \nu} F^{a \mu \nu } + \sum_{i=1}^{n_f} \bar{q}_i i \slashed{D} q_i  
\nonumber \\
&+& \int  d^3 \bm{r} \; \text{Tr}\left\lbrace {\rm S}^{\dagger} \left( i\partial_0 -h_s \right) {\rm S} + {\rm O}^{\dagger } \left( i D_0 -h_o \right) {\rm O} \right\rbrace 
\nonumber \\
& &\hspace{1.5cm} + \text{Tr} \left\lbrace  {\rm O}^{\dagger } \bm{r} \cdot g\bm{E} {\rm S} + {\rm S}^{\dagger} \bm{r} \cdot \bm{E} {\rm O} \right\rbrace  + \dots  \,,
\end{eqnarray}
where $\bm{r}$ is the heavy quark-antiquark distance vector, 
${\rm S} = S \mathbbm{1}_c/\sqrt{N_c}$ and ${\rm O} = O^a T^a /\sqrt{T_F}$ are the heavy quark-antiquark color-singlet and color-octet fields respectively,
$q_i$ are $n_f$ light quark fields taken massless, 
$N_c$ is the number of colors, $T_F=1/2$, and traces are understood over color and spin indices. 
We have taken the matching coefficients at leading order. 
The dots stand for higher-order terms in the multiple expansion and for octet-octet transitions that we do not need in the following. 
The singlet and octet Hamiltonians read
\begin{equation}
h_{s,o}=\frac{\bm{p}^2}{m} + V^{(0)}_{s,o} + \dots  \, ,
\end{equation}
where $\bm{p} = -i {\bm\nabla}_{\bm{r}}$ and the dots stand for higher-order terms in the $1/m$ expansion. 
The singlet and octet static potentials are at leading order in $\alpha_{\rm s}$:
$V^{(0)}_s=-C_F \alpha_{\rm s}/r$ and $V^{(0)}_o=\alpha_{\rm s}/(2N_c \, r)$ respectively; 
$C_F=(N_c^2-1)/(2 N_c)$ is the Casimir of the fundamental representation of SU$(N_c)$.

The computations that we are going to perform in this and in the next section share similarities 
with the ones done for quarkonium in a hot wind in the same temperature regime~\cite{Escobedo:2011ie,Escobedo:2013tca}. 
In both cases we are dealing with a problem in which the distribution of particles in the medium has a preferred direction.

Thermal contributions to the real and imaginary parts of the heavy-quarkonium spectrum come from considering self-energy diagrams in pNRQCD 
and integrating them over momentum regions scaling respectively like the temperature and the binding energy.
Integrating over the momentum region scaling like the temperature amounts at matching pNRQCD to another effective field theory, 
dubbed pNRQCD$_{\hbox{\tiny HTL}}$ in~\cite{Brambilla:2008cx,Brambilla:2010vq}, 
where only modes with energy and momentum smaller than $\pi T$ are dynamical.
Thermal contributions are then encoded in the color-singlet potential of pNRQCD$_{\hbox{\tiny HTL}}$. 
We will consider integrating over the momentum region scaling like the binding energy in the next section.

\begin{figure}[ht]
\includegraphics[scale=0.65]{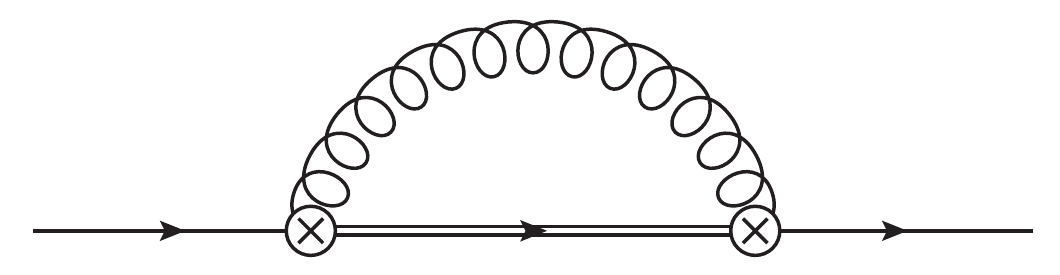}
\caption{\label{singlet_octet} 
Color-singlet self-energy diagram in pNRQCD. Single lines stand for
quark-antiquark color-singlet propagators, double lines for color-octet propagators, curly lines for gluons 
and a circle with a cross for a chromoelectric dipole vertex.}
\end{figure}

The leading thermal contribution to the color-singlet potential comes from the self-energy diagram in Fig.~\ref{singlet_octet}, 
where the loop momentum is set to be $q \sim \pi T$.  
By using the vertices and propagators of the pNRQCD Lagrangian we obtain
\begin{eqnarray}
&&\langle  \Omega  |{\rm T} \, S(t,\mathbf{r}, \mathbf{R}) \, S^{\dagger}(0,\mathbf{0}, \mathbf{0}) | \Omega \rangle = -4 \pi \alpha_{\rm s} C_F  
\nonumber \\
&&\times 
\int_P  e^{-iP_0 t +i \bm{P} \cdot \mathbf{R}} \langle \mathbf{r} |\frac{i}{P_0 - h_s + i\epsilon} \, r_i I_{ij} r_j \, \frac{i}{P_0 - h_s + i\epsilon}| \mathbf{0} \rangle ,
\nonumber 
\\
\label{correlator}
\end{eqnarray} 
where T stands for time ordering, $\int_P \equiv \int d^4P /(2\pi)^4$, $P^{\mu}=(P^0,\bm{P})$ and $| \Omega \rangle$ is the ground state of the theory. 
The thermal part of the self-energy loop integral, $I_{ij}$, is given by 
\begin{equation}
I_{ij}=\int_q  \frac{i(q_0)^2 \, 2 \pi \, \delta(q^2)}{P_0-q_0-h_o+i\epsilon} \left( \delta_{ij} 
- \frac{q_i q_j}{|\mathbf{q}|^2}\right) \,  f^{\rm B}(\mathbf{q},\xi) \, .
\label{Int_a}
\end{equation}
We have to separate terms that go into the wave-function renormalization from those that go into the color-singlet potential of pNRQCD$_{\hbox{\tiny HTL}}$. 
To this end we rewrite $P^0 - h_o = P^0 - h_s - \Delta V$, where $\Delta V=(h_o-h_s) = (N_c \alpha_{\rm s})/(2 r) + \dots$, 
and, due to the condition $q \sim \pi T$ that sets the loop momentum to be much larger than the energy of the heavy quark-antiquark pair,  
we expand the octet propagator in \eqref{Int_a}. 
After dropping terms that go into the wave-function renormalization, 
the part of $r_i  I_{ij}  r_j$ that contributes to the color-singlet potential of pNRQCD$_{\hbox{\tiny HTL}}$ reads
\begin{eqnarray}
&& \left. r_i  I_{ij}  r_j \right|_{q\sim \pi T}^{{\rm contr.\,to\,}V_s} = i\left( \frac{2}{m} + \Delta V \, r^2\right) \int_q  \, 2 \pi \, \delta(q^2) f^{\rm B}(\mathbf{q},\xi) 
\nonumber \\
&& \hspace{2.3cm} -i (\Delta V r_i r_j)\, \int_q  2 \pi \, \delta(q^2)  \frac{q_i q_j}{|\mathbf{q}|^2} f^{\rm B}(\mathbf{q},\xi) \, .
\label{integralmatch}
\end{eqnarray}

To match onto pNRQCD$_{\rm{{HTL}}}$ we compute the correlator 
$\langle  \Omega  |{\rm T} \, S(t,\mathbf{r}, \mathbf{R}) \, S^{\dagger} (0,\mathbf{0}, \mathbf{0}) | \Omega \rangle$ in pNRQCD$_{\hbox{\tiny HTL}}$ 
and require this expression to be equal to \eqref{correlator}.
The color-singlet potential of pNRQCD$_{\hbox{\tiny HTL}}$ turns out to be the same as in pNRQCD plus a thermal correction $\delta V_s$ that reads
\begin{equation}
\delta V_s = -i 4 \pi \alpha_{\rm s} C_F \left. r_i  I_{ij}  r_j \right|_{q\sim \pi T}^{{\rm contr.\,to\,}V_s} \, .
\label{match1}
\end{equation}
The integral can be easily evaluated and the final result for the anisotropic potential at finite temperature is 
\begin{eqnarray}
\delta V_s &=&  \frac{2\pi \alpha_{\rm s} C_F T^2}{3 m} \mathcal{F}_1(\xi) 
+ \frac{\pi \alpha_{\rm s}^2 C_F N_c T^2 r}{12} \mathcal{F}_2(\xi) 
\nonumber \\
&+& \frac{\pi \alpha_{\rm s}^2 C_F N_c T^2 (\bm{r}\cdot \bm{n})^2}{12 r} \mathcal{F}_3(\xi) \, ,
\label{resultVs}
\end{eqnarray}  
where the definitions of the functions embedding the anisotropy parameter are 
\begin{eqnarray}
&&\mathcal{F}_1(\xi) = N(\xi)\frac{\arctan \sqrt{\xi}}{\sqrt{\xi}}  ,
\label{f1}
\\
&&\mathcal{F}_2(\xi)= N(\xi)\left( \frac{\arctan \sqrt{\xi}}{\sqrt{\xi}} + \frac{1}{\xi} -\frac{\arctan \sqrt{\xi}}{\xi \sqrt{\xi}}\right) ,
\label{f2}
\\
&&\mathcal{F}_3(\xi)= N(\xi)\left( \frac{\arctan \sqrt{\xi}}{\sqrt{\xi}} - \frac{3}{\xi} +\frac{3 \arctan \sqrt{\xi}}{\xi \sqrt{\xi}} \right) .
\label{f3}
\end{eqnarray}

We comment briefly about the result: 
first, at this order no imaginary part, and hence no thermal width, arises; 
second, for $\xi \to 0$ the result in \eqref{resultVs} agrees with the isotropic case derived in~\cite{Brambilla:2010vq}. 
Finally, we notice that the term in the second line in \eqref{resultVs} is of order $\xi$ when expanding for a
small anisotropy parameter, signaling that its origin is entirely due to the breaking of the spherical symmetry of the parton momentum distribution.

\section{Thermal corrections to the spectrum}
\label{sec_2}
In our setting the next relevant scale after the temperature is the quarkonium binding energy. 
The process we are looking at is again a singlet-to-octet transition, 
however with energy and momenta scaling like $m\alpha_{\rm s}^2$ rather than~$\pi T$. 
This contribution is not part of the potential but comes as a low-energy correction to the spectrum of pNRQCD$_{\rm{{HTL}}}$. 
It may be computed at leading order from the one-loop diagram in Fig.~\ref{singlet_octet},  
where now, however, the typical loop momentum is selected to be of order $m\alpha_{\rm s}^2$. 
To ensure that we are computing only contributions from the momentum region $q \sim m\alpha_{\rm s}^2 \ll \pi T$, 
we need to expand the anisotropic distribution function 
\begin{equation}
f^{\rm B}(\mathbf{q},\xi)=\left( e^{\frac{|\mathbf{q}|}{T}\sqrt{1+\xi \lambda^2}} -1\right)^{-1}  \approx \frac{T}{|\mathbf{q}| \sqrt{1+\xi \lambda^2}}\, ,
\end{equation}
where $\lambda=\bm{q} \cdot \bm{n} /|\bm{q}|$ is the cosine of the angle between the gluon momentum and the anisotropy direction.
We keep only the leading term in the $|\bm{q}|/T$ expansion.  
Differently from the calculation in Sec.~\ref{sec_1}, we cannot expand the octet propagator.
Then the contribution from the momentum region $q \sim m\alpha_{\rm s}^2$ to the self-energy diagram in Fig.~\ref{singlet_octet} reads
\begin{equation}
\delta \Sigma= -i4 \pi \alpha_{\rm s} C_F r_i \left. I_{ij} r_j \right|_{q\sim m\alpha_{\rm s}^2}\, ,
\label{self_E}
\end{equation}
where 
\begin{eqnarray}
&&\left. r_i I_{ij} r_j\right|_{q\sim m\alpha_{\rm s}^2} = T r_i\int_q \frac{i(q_0)^2 \,  2 \pi \,\delta(q^2)}{P_0-q_0-h_o+i\epsilon} \left( \delta_{ij} - \frac{q_i q_j}{|\mathbf{q}|^2}\right) 
\nonumber
\\
&& \hspace{4 cm} \times \frac{r_j}{|\mathbf{q}| \sqrt{1+\xi \lambda^2}}  \,.
\label{secInt}
\end{eqnarray}

The integral \eqref{secInt} has a vanishing imaginary part. This means that there is no contribution coming from $\delta \Sigma$, 
as defined in \eqref{self_E}, to the real part of the spectrum. Hence, the thermal shift in the binding energy 
is entirely due to the shift in the singlet potential, $\delta V_s$, computed previously in \eqref{resultVs}.
We can write it as
\begin{equation}
\delta E_{{\rm{bind}}}= \langle n \, l \, m | \delta V_{s} | n \, l \, m\rangle  \,,
\end{equation} 
where $| n \, l \, m\rangle$ are eigenstates of the singlet Hamiltonian $h_s$, with quantum numbers $n$, $l$ (orbital angular momentum)
and $m$ (orbital angular momentum along the $z$ direction).
Since, according to our hierarchy of energy scales, the potential entering $h_s$ is the Coulomb potential, 
the states $| n \, l \, m\rangle$ are just Coulombic bound states. 
At leading accuracy, $\delta E_{{\rm{bind}}}$ then reads
\begin{eqnarray}
&& \delta  E_{{\rm{bind}}} = \frac{2\pi \alpha_{\rm s} C_F T^2}{3 m} \mathcal{F}_1(\xi)  
\nonumber\\
&& \hspace{1.1cm}
+\frac{\pi \alpha_{\rm s} N_c T^2}{12m} \left[ 3n^2 -l(l + 1) \right] \left( \mathcal{F}_2(\xi)  + \frac{\mathcal{F}_3(\xi)}{3}  
\right.
\nonumber\\
&& \hspace{3cm}
\left. + \frac{2}{3} \mathcal{F}_3(\xi) \, C^{l \, 0}_{2 \, l \, 0 \, 0} \, C^{l \, m}_{2 \, l \, 0 \, m} \right) ,
\label{resultani1}
\end{eqnarray}
where the Clebsch--Gordan coefficients are understood with the notation 
$C^{JM}_{j_1 j_2 m_1 m_2}$ ($C^{JM}_{j_1 j_2 m_1 m_2} = 0$ if $J > j_1+j_2$ or $J < |j_1-j_2|$).

\begin{table}[ht]
\begin{ruledtabular}
\begin{tabular}{c|c|c|c|c|c}
$\xi$ & $\mathcal{F}_1(\xi)$  & $\mathcal{F}_2(\xi)$ & $\mathcal{F}_3(\xi)$ & $\mathcal{G}_1(\xi)$  &  $\mathcal{G}_2(\xi)$ \\
\hline
$0.1$  & 1.016  & 1.346 & 0.026 & 1.032 & 0.009
\\
\hline
$0.3$ & 1.043 & 1.367 & 0.072 & 1.089 & 0.026
\\
\hline
$0.5$ & 1.067 & 1.383  & 0.114 & 1.141 & 0.041
\\
\hline
$1$ & 1.110  & 1.414 & 0.200 & 1.246 & 0.077
\end{tabular}
\end{ruledtabular}
\caption{\label{table_functions} 
The anisotropy functions defined in \eqref{f1}-\eqref{f3}, \eqref{g1} and \eqref{g2} for some values of $\xi$. }
\end{table}

The integral \eqref{secInt} has a non-vanishing real part that contributes to the imaginary part of  $\delta \Sigma$.
The imaginary part of the self energy gives rise to a thermal width:
\begin{eqnarray}
\Gamma &=& - 2\langle n  l m | {\rm{Im}} \, (\delta \Sigma) | n  l m \rangle 
\nonumber\\
&=& 8 \pi^2 \alpha_{\rm s} C_F T \,
\langle  n  l m | r_i \,  \int_q \frac{\delta(E_n + q_0 - h_o) \, q^2_0 }{|\mathbf{q}| \sqrt{1+\xi \lambda^2}} 
\nonumber \\
&& \hspace{1.3cm} \times 
\left( \delta_{ij} -\frac{q_i q_j}{|\mathbf{q}|^2} \right)  (2\pi)\delta(q^2) \, r_j  | n  l m \rangle \,,
\label{gammastart}
\end{eqnarray}
where $E_n = -m (C_F \alpha_{\rm s})^2/(4n^2)$ is the energy of the bound state. 
The final result reads
\begin{eqnarray}
&& \Gamma =  \frac{4}{3}\alpha_{\rm s}^3T\left(\frac{C_FN_c^2}{4} + \frac{C_F^2N_c}{n^2} + \frac{C_F^3}{n^2}\right)\mathcal{G}_1(\xi)  
\nonumber\\
&& \hspace{0.2cm} 
+ \alpha_{\rm s}^3T\left(\frac{C_FN_c^2}{4} - \frac{C_F^2N_c}{2n^2} + \frac{C_F^3}{n^2}\right)
\mathcal{G}_2(\xi) C^{l \, 0}_{2 \, l \, 0 \, 0} \, C^{l \, m}_{2 \, l \, 0 \, m} \, , 
\nonumber\\
\label{resultgamma}
\end{eqnarray}   
where the anisotropy functions are in this case 
\begin{eqnarray}
&&\!\!\!\!\!\!\!\mathcal{G}_1(\xi) = N(\xi) \frac{\text{arcsinh} \left( \sqrt{\xi}\right)}{\sqrt{\xi}}  ,
\label{g1}
\\
&&\!\!\!\!\!\!\!\mathcal{G}_2(\xi) = N(\xi)\frac{\left(1+ 2\xi/3 \right)\text{arcsinh} \left( \sqrt{\xi}\right) -\sqrt{\xi(1+\xi)} }{\sqrt{\xi^3}}  .
\label{g2}
\end{eqnarray}
The appearance of a thermal width follows from the fact that the singlet-to-octet transition becomes 
a real process if the emitted gluon has an energy of the order of the binding energy.

The limit $\xi \to 0$ corresponds to the isotropic case.
For $\xi \to 0$, we have that $\mathcal{F}_1(\xi) \to 1$,  $\mathcal{F}_2(\xi) \to 4/3$, $\mathcal{G}_1(\xi) \to 1$, 
whereas both $\mathcal{F}_3(\xi)$ and $\mathcal{G}_2(\xi)$ vanish linearly in $\xi$. 
In this limit both the binding energy \eqref{resultani1} and the thermal width \eqref{resultgamma} 
reduce to previously known expressions found in~\cite{Brambilla:2010vq}.  
In Tab.~\ref{table_functions} we show some benchmark values of the anisotropy functions.

\begin{figure}[ht]
\includegraphics[scale=0.61]{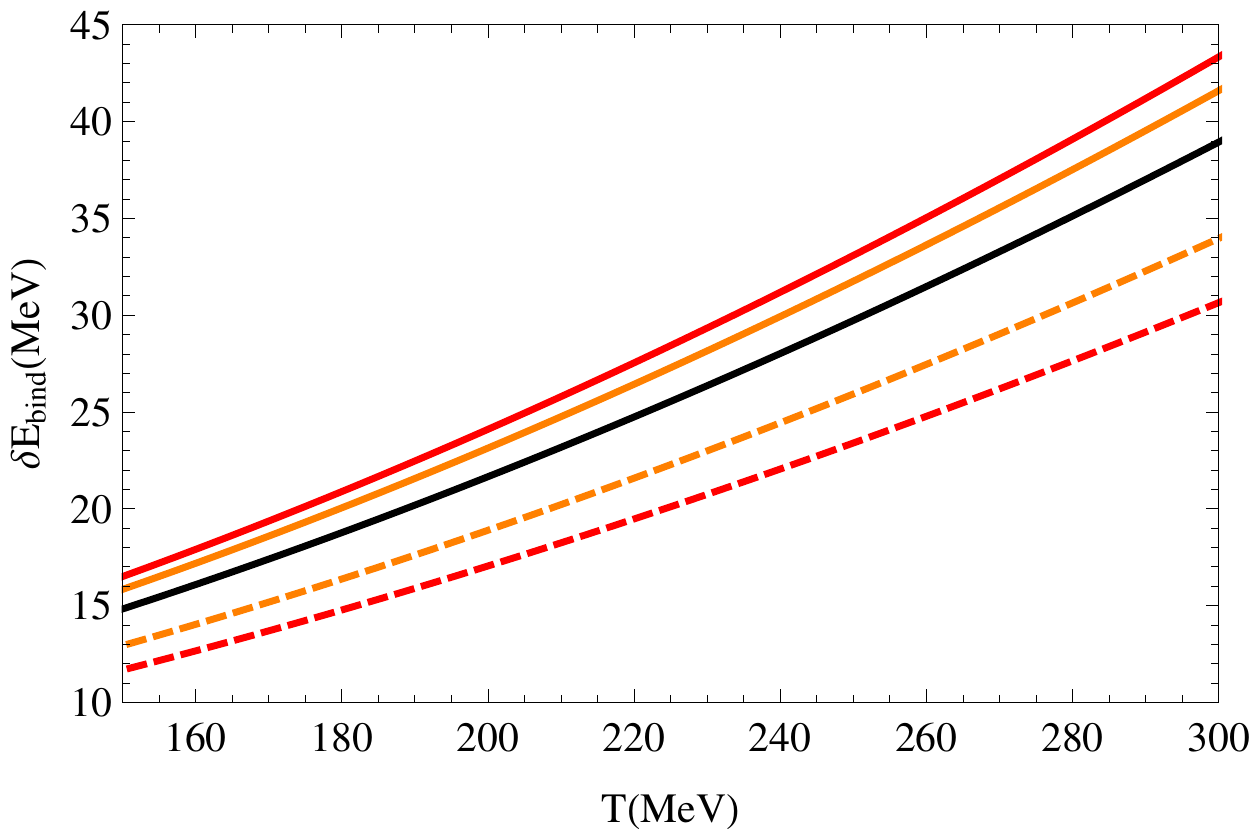}
\caption{\label{bindl0} 
(Color Online) Binding-energy shift of a $1S$ ($n=1,l=0$) bottomonium state according to \eqref{resultani1}. 
We show the binding-energy shift for the isotropic case, black solid line, 
and for two different values of the anisotropy parameter $\xi=0.5$ and $\xi=1$ in orange and red solid (dashed) lines 
respectively when the normalization factor is $N(\xi)=\sqrt{1+\xi}$ ($N(\xi)=1$). 
For all the figures (here and in the following) we have taken $\alpha_{\rm s}(2 \pi T)$ and considered it running at one loop with three quark flavours.
The bottom-quark mass has been chosen to be half of the $\Upsilon(1S)$ mass, i.e., 4730~MeV.}
\end{figure}

\begin{figure}[ht]
\includegraphics[scale=0.61]{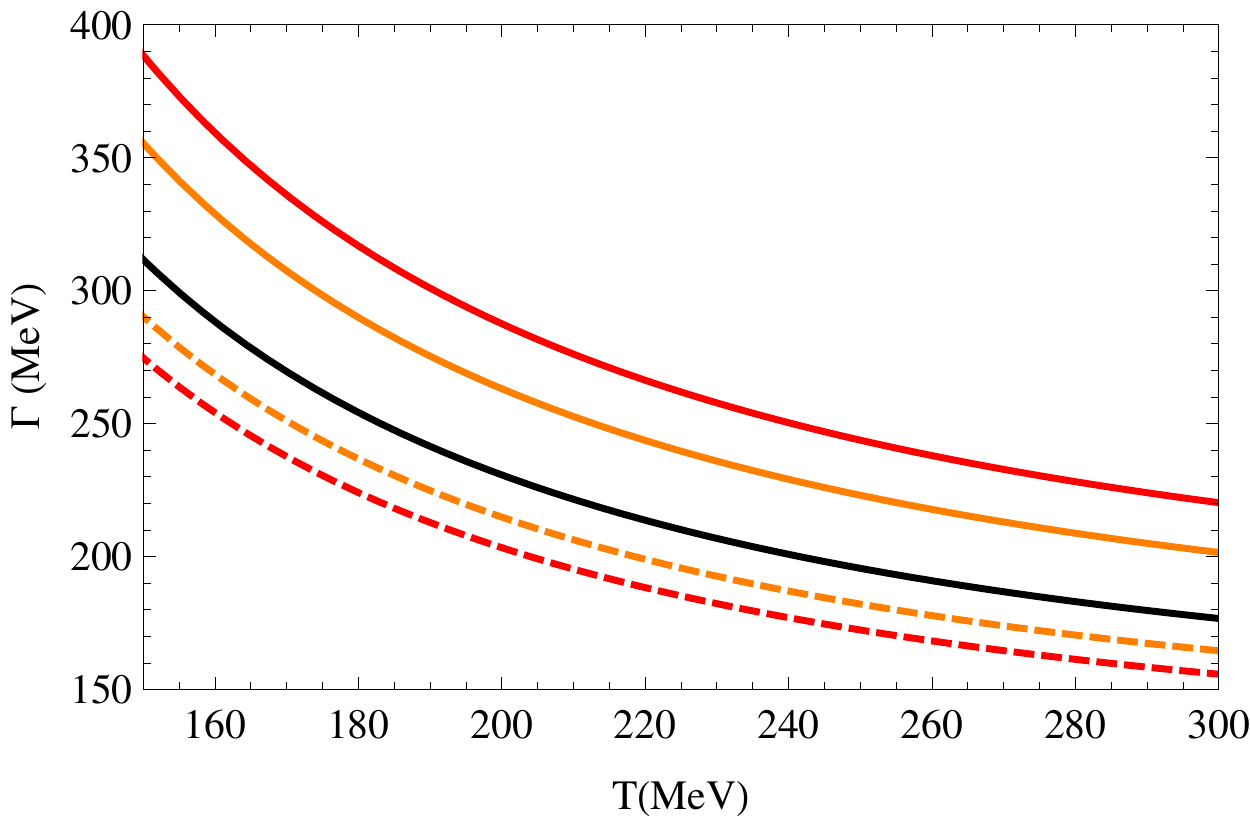}
\caption{\label{gammal0} 
(Color Online) Thermal width of a $1S$ ($n=1,l=0$) bottomonium state according to \eqref{resultgamma}. 
The different curves are defined as in Fig.~\ref{bindl0}.}
\end{figure}

\section{Conclusion and discussion}
\label{sec_3}
In an early stage, heavy-ion collisions are characterized by parton momentum anisotropies.
Accordingly the evolution of the fireball is described in terms of viscous and anisotropic hydrodynamical models. 
Due to the fact that hard probes, like heavy quarkonia, get formed in such an early stage of the heavy ion-collisions 
and experience the medium until late times, their dynamics has to account for an anisotropic momentum distribution of the QGP constituents. 
In this paper, we have derived for the hierarchy of scales~\eqref{hiera} and at leading order the real and imaginary thermal parts of the quarkonium spectrum in an anisotropic QGP.
The imaginary part originates from the quarkonium gluo-dissociation in the medium. 
Our result complements previous studies for an anisotropic plasma where the real and imaginary part 
of the quark-antiquark potential were obtained for a temperature scale larger than the inverse radius of the bound state. 
In so doing we extend the knowledge of a weakly-coupled quarkonium to temperature
ranges that may be reached during the QGP evolution at present day colliders.

\begin{figure}[ht]
\includegraphics[scale=0.62]{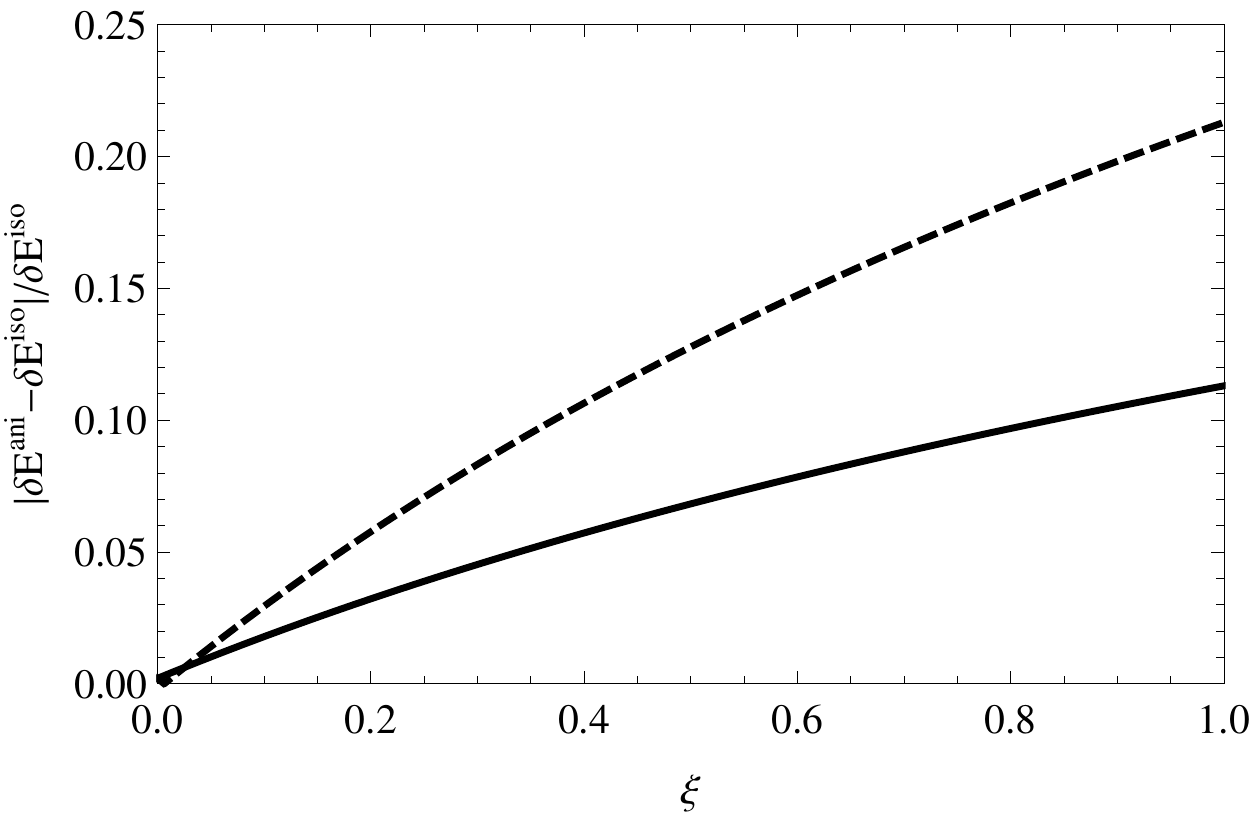}
\hspace{0.2 cm}
\includegraphics[scale=0.62]{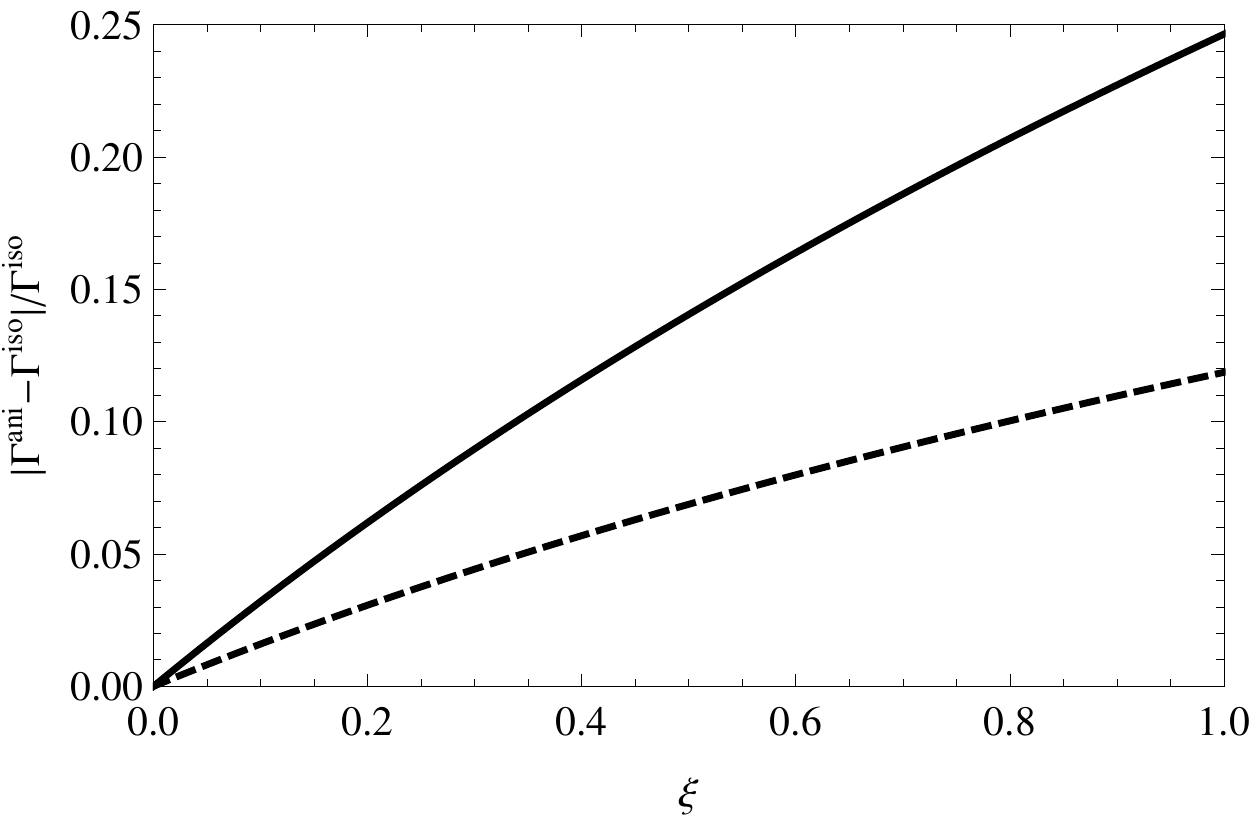}
\caption{\label{delta_ani} 
Relative change in the binding energy (upper plot) and thermal width (lower plot) due to the presence of a momentum anisotropy.
$\delta E^{\hbox{\tiny ani}}$ is the binding-energy shift in \eqref{resultani1} evaluated for $\xi \neq 0$, 
whereas $\delta E^{\hbox{\tiny \rm iso}}$ is the binding-energy shift in \eqref{resultani1} evaluated at $\xi=0$, for a $1S$ bottomonium state. 
In a similar way we have defined the thermal widths, $\Gamma^{\hbox{\tiny ani}}$ and $\Gamma^{\hbox{\tiny \rm iso}}$, taken from \eqref{resultgamma}.
For solid (dashed) lines the normalization has been taken $N(\xi)=\sqrt{1+\xi}$ ($N(\xi)=1$).
}
\end{figure}

\begin{figure}[ht]
\includegraphics[scale=0.64]{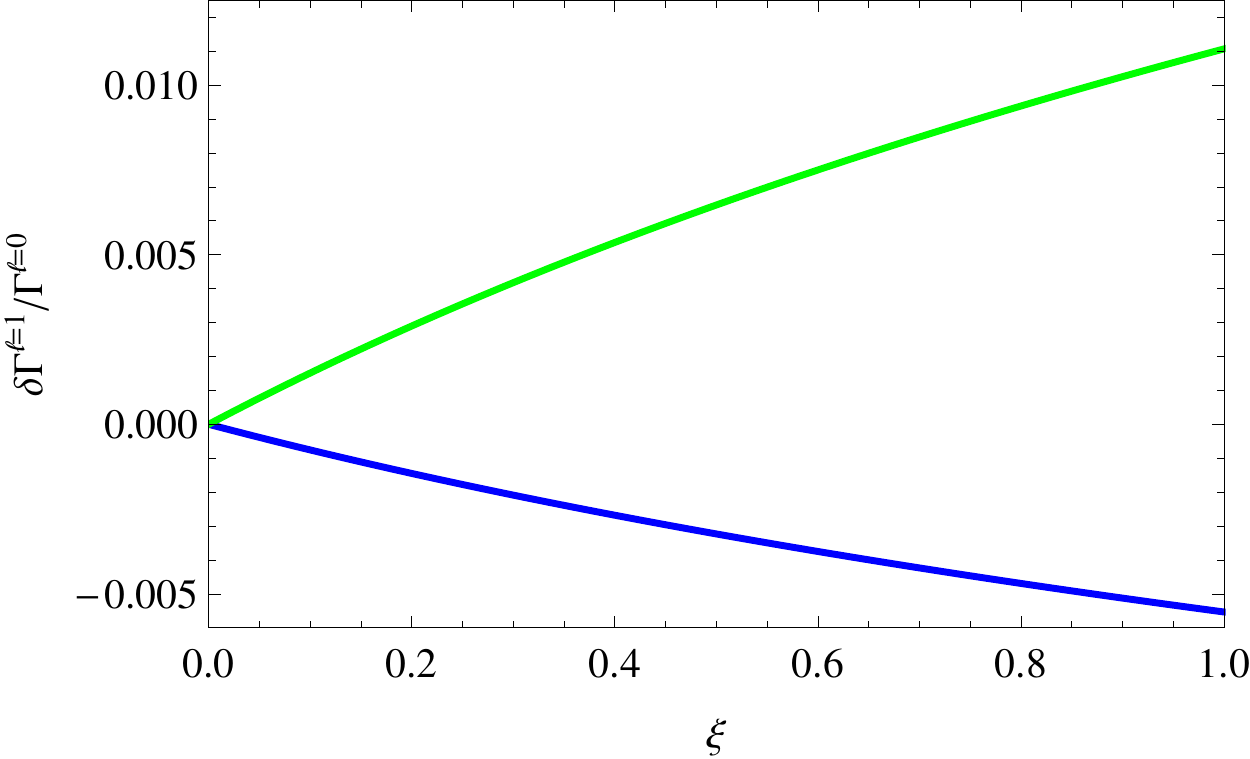}
\caption{\label{fig_gammal_l} 
(Color Online) Ratio of the differences between the thermal corrections to the widths of $1P$ ($n=2,l=1$) 
and $2S$ ($n=2, l=0$) bottomonium states and the $2S$-state thermal width, respectively second and first line in \eqref{resultgamma}, 
as a function of $\xi$. The blue (lower) line refers to the $m=\pm 1$ states whereas the green (upper) line to the $m=0$ one.}
\end{figure}

The real thermal part of the spectrum comes from thermal corrections to the potential defined in the context of pNRQCD$_{\rm{{HTL}}}$.
They are encoded in the self-energy diagram of Fig.~\ref{singlet_octet} evaluated at the temperature scale. 
The result is given in \eqref{resultVs}. 
Thermal corrections to the potential are proportional to the square of the temperature 
and, as discussed elsewhere, do not show Debye screening~\cite{Brambilla:2008cx, Brambilla:2010vq}. 
The corresponding expectation value provides the real part of the thermal corrections to the binding energy.
They are given in \eqref{resultani1}.  
In Fig.~\ref{bindl0} we show the binding-energy shift for a $1S$ bottomonium state in the isotropic case,
$\xi=0$, and in the case of a finite momentum anisotropy, $\xi=0.5$ and $\xi=1$.
We see that the impact of an anisotropic plasma crucially depends on the normalization factor, 
either $N(\xi)=1$ or $N(\xi)=\sqrt{1+\xi}$, respectively shown in dashed and solid lines.
For $N(\xi)=1$ the anisotropy reduces the thermal correction to the binding energy, whereas for $N(\xi)=\sqrt{1+\xi}$ it increases it.

The computation of the spectrum in pNRQCD$_{\rm{{HTL}}}$ leads also to an imaginary part 
coming from the self-energy diagram of Fig.~\ref{singlet_octet} evaluated at the binding-energy scale. 
The imaginary part may be understood as a thermal width, whose explicit expression is in \eqref{resultgamma}.  
In Fig.~\ref{gammal0} we show the thermal width for a $1S$ bottomonium state in the isotropic case, $\xi=0$, 
and in the case of a finite momentum anisotropy, $\xi=0.5$ and $\xi=1$. 
Also here the size and sign of the thermal corrections strongly depend on the normalization factor,
either $N(\xi)=1$ or $N(\xi)=\sqrt{1+\xi}$, respectively shown in dashed and solid lines. 
Although the dependence on the anisotropy is qualitatively similar in the binding energy and thermal width,  
we find that the effect of the anisotropy is more important for the binding energy with respect to the thermal width 
when $N(\xi)=1$ (see dashed lines in Fig.~\ref{delta_ani}), 
whereas the opposite is true when $N(\xi)=\sqrt{1+\xi}$ (see solid lines in Fig.~\ref{delta_ani}).

Finally, we comment on the effect of an anisotropic QGP on the bound-state polarization. 
In Fig.~\ref{fig_gammal_l} we show the differences between the thermal corrections to the widths of $1P$ and $2S$ bottomonium states. 
For $\xi \le 1$ such differences are typically of the order of few per mill (at most 1\%) with respect to the corresponding $2S$ state thermal width.
This suppression is due to various effects: 
the ratio between the anisotropy functions $\mathcal{G}_2$ and $\mathcal{G}_1$, see the benchmark values in Tab.~\ref{table_functions}, 
the combination involving the color factors $N_c$ and $C_F$, and the Clebsch--Gordan coefficients. 
We conclude that for small anisotropies the effect of an anisotropic QGP on the bound-state polarization is tiny and possibly phenomenologically irrelevant.

\section*{Acknowledgements}
The work of S.B. was partly supported by the Swiss National Science Foundation (SNF) under grant 200020-168988.
N.B. and A.V. acknowledge support from the DFG grant BR 4058/1-2 ``Effective field theories for hard probes of hot plasma" 
and the DFG cluster of excellence ``Origin and structure of the universe" (www.universe-cluster.de).
The work of M.A.E. was supported by the European Research Council under the Advanced Investigator Grant ERC-AD-267258
and by the Academy of Finland, project~303756.

\bibliography{biblio.bib}

\end{document}